\documentclass[preprint,3p,times]{elsarticle}

\usepackage{amssymb}
\usepackage{amsthm}
\usepackage{amsmath}
\usepackage{epsfig}
\usepackage{subfigure}
\usepackage[usenames]{color}
\usepackage{algorithm}
\usepackage{algorithmic}
\usepackage{nicefrac}
\usepackage{caption}
\epsfverbosetrue
\graphicspath{{fig/}{../fig/}}

\newcommand{\mb}[1]{\ensuremath{\mathbf{#1}}}

\journal{Powder Technology}

\begin{document}
 \setlength{\parindent}{0.0ex}
 \setcounter{secnumdepth}{4}
 \setcounter{tocdepth}{4}
\begin{frontmatter}







\title{Modeling and Characterization of Cohesion in Fine Metal Powders with a Focus on Additive Manufacturing Process Simulations}



\author[mechanosynth,lnm]{Christoph Meier\corref{cor1}}
\ead{meier@lnm.mw.tum.de}
\author[mechanosynth,lnm]{Reimar Weissbach}
\author[mechanosynth,lincoln]{Johannes Weinberg}
\author[lnm]{Wolfgang A. Wall}
\author[mechanosynth]{A. John Hart\corref{cor1}}
\ead{ajhart@mit.edu}

\address[mechanosynth]{Mechanosynthesis Group, Department of Mechanical Engineering, Massachusetts Institute of Technology, 77 Massachusetts Avenue, Cambridge, 02139, MA, USA}
\address[lnm]{Institute for Computational Mechanics, Technical University of Munich, Boltzmannstrasse 15, 85748 Garching b. M{\"u}nchen, Germany}
\address[lincoln]{M.I.T. Lincoln Laboratory, 244 Wood Street, Lexington, 02420, MA, USA}

\cortext[cor1]{Corresponding authors}


\begin{abstract}
The cohesive interactions between fine metal powder particles crucially influence their flow behavior, which is in turn important to many powder-based manufacturing processes including emerging methods for powder-based metal additive manufacturing (AM). The present work proposes a novel modeling and characterization approach for micron-scale metal powders, with a special focus on characteristics of importance to powder-bed AM. The model is based on the discrete element method (DEM), and the considered particle-to-particle and particle-to-wall interactions involve frictional contact, rolling resistance and cohesive forces. Special emphasis lies on the modeling of cohesion. The proposed adhesion force law is defined by the pull-off force resulting from the surface energy of powder particles in combination with a van-der-Waals force curve regularization. The model is applied to predict the angle of repose (AOR) of exemplary spherical Ti-6Al-4V powders, and the surface energy value underlying the adhesion force law is calibrated by fitting the corresponding angle of repose values from numerical and experimental funnel tests. To the best of the authors' knowledge, this is the first work providing an experimental estimate for the effective surface energy of the considered class of metal powders. By this approach, an effective surface energy of $0.1mJ/m^2$ is found for the investigated Ti-6Al-4V powder. This value is considerably lower than typical experimental values for flat metal contact surfaces in the range of $30-50 mJ/m^2$, indicating the crucial influence of factors such as surface roughness and potential chemical surface contamination / oxidation on fine metal powders. More importantly, the present study demonstrates that a neglect of the related cohesive forces leads to a drastical underestimation of the AOR and, consequently, to an insufficient representation of the bulk powder behavior.
\end{abstract}
\end{frontmatter}

%
%
\section{Introduction}
\label{sec:intro}
%
%

There is a variety of technical processes relying on metal powder as raw material, with powder-bed fusion additive manufacturing certainly representing the most prominent class of these processes. Among the manifold of existing additive manufacturing (AM) processes, selective laser melting (SLM) of metals has attracted much scientific attention in recent years since it offers near-net-shape production of virtually limitless geometries, and eventual potential for pointwise control of microstructure and mechanical properties~\cite{Gibson2010,Stucker2011,Thijs2010}. On the other hand, the overall SLM process is complex and governed by a variety of (competing) physical mechanisms. A sub-optimal choice of process parameters might lead to deteriorated material properties or even to failure of the part already during the manufacturing process~\cite{Das2003,Gong2014,Kruth2007,Kruth2005,Sames2016}. The complexity and sensitivity of the process leads to an inevitable demand for experimental and numerical studies in order to gain an in-depth understanding of the underlying physical mechanisms and ultimately to optimize the process and the properties of the final part.\\

Depending on the lengths scales under consideration, existing experimental and numerical studies can typically be classified (see e.g. ~\cite{Meier2018}) in macroscopic approaches analyzing thermo-mechanical effects such as residual stresses or dimensional warping on part level~\cite{Denlinger2014,Gusarov2007,Hodge2016,Pal2014,Riedlbauer2016,Zaeh2010}, mesoscopic approaches commonly investigating melt pool fluid dynamics within single laser beam tracks by resolving length scales in the range of the powder particle size~\cite{Ammer2014,Khairallah2016,Korner2011,Lee2015,Megahed2016,Qiu2015} as well as microscopic approaches studying the evolution of the metallurgical microstructure during the process~\cite{Dorr2010,Gong2015,Rai2016,Zhang2013}. Most of these studies intend to optimize the SLM process by relating the observed process outcomes with input parameters such as laser beam power and velocity, powder layer thickness, hatch spacing or scanning strategy. There are also some - but considerably less - contributions that have studied the influence of the powder feedstock on the process outcome~\cite{Boley2015,Boley2016,Gusarov2008,Gusarov2005,Herbert2016,Tan2017,Whiting2016}. However, only very few of these works have studied the actual interplay of powder particle properties (mechanical, thermal, optical and chemical properties on particle surfaces), bulk powder properties (morphology, granulometry and resulting flowability) as well as resulting powder layer characteristics (packing density, surface uniformity as well as effective thermal and mechanical properties) during the powder recoating process applied between two subsequent material layers in metal additive manufacturing. Among others, this deficit has been reflected by a recent priority meeting on metal powder modeling of political, scientific and industrial leaders in the field of AM technologies~\cite{King2017}.\\

The few existing studies of powder recoating processes in metal AM comprise the pioneering works of Herbold et al.~\cite{Herbold2015}, Mindt et al.~\cite{Mindt2016}, Haeri et al.~\cite{Haeri2017} and Gunasegaram et al.~\cite{Gunasegaram2017}. All of these approaches employed a model based on the discrete element method (DEM) typically accounting for (visco-) elastic normal contact, sliding friction as well as rolling friction interaction between spherical particles. The mentioned works demonstrated the general applicability of DEM-based models and already gained valuable insights into physical mechanism underlying the powder re-coating process. However, none of these pioneering modeling approaches have considered powder cohesiveness so far, allthough its importance has already been argued in~\cite{Herbold2015}.\\

It is well-known that bulk powder cohesiveness increases with decreasing particle size, which is a direct consequence of the cubic and linear scaling with particle size typically observed for volume forces (e.g. gravity forces) and adhesive forces (e.g. van-der-Waals forces associated with the particle's surface energy), respectively~\cite{Tan2017,Walton2008}. When considering typical surface energy values measured for metals, it can be concluded that cohesive forces might crucially influence the bulk powder behavior of micron-scale metal powders typically applied in AM processes~\cite{Herbold2015}. While there are some existing approaches incorporating cohesion in DEM-based powder coating / packing models, to the best of the authors' knowledge, none of these contributions has considered metallic powder materials~\cite{Parteli2016,Parteli2014,Xiang2016}. It is questionable if the results derived by these studies can directly be transferred to metal additive manufacturing, since the underlying physical mechanism as well as the typical magnitudes of cohesive forces are crucially different for (conductive) metallic and non-metallic powders. Moreover, to the best of the authors' knowledge, there is currently no experimental data available in the literature that would enable a quantification of the effective surface energy, and thus of the cohesive forces, prevalent in these powders. However, an experimental characterization of the powders at hand is important since surface energy values for a given material combination can easily vary by several orders of magnitude as consequence of surface roughness and potential surface contamination / oxidation.\\

The present work closes this gap by proposing a powder model based on the discrete element method (DEM) with particle-to-particle and particle-to-wall interactions involving frictional contact, rolling resistance and cohesive forces. Special emphasis lies on the modeling of cohesive effects. The proposed adhesion force law is defined by the pull-off force associated with the surface energy of powder particles in combination with a van-der-Waals force curve regularization. For the first time, the magnitude of the effective surface energy and of the resulting cohesive forces will be quantified for the considered class of metal powders by fitting angle of repose (AOR) values resulting from experimental and numerical funnel tests. As a result, it will be found that the effective surface energy determined for the investigated micron-scale metal powder is lower than typical values measured for sheet metal~\cite{Herbold2015,Santos2004}. More importantly, it is shown that a neglect of the resulting cohesive forces leads to a drastical underestimation of the AOR and, consequently, to an insufficient description of the bulk powder behavior in potential powder recoating simulations based on such a simplified model. Eventually, the present study demonstrates that the resulting angle of repose is more sensitive concerning the magnitude of surface energy as compared to the magnitude of other powder material parameters such as powder particle stiffness, friction coefficient or coefficient of restitution. The proposed modeling and characterization approach is carried on with having powder-bed fusion additive manufacturing processes in mind, but is generally valid for a broad range of applications involving micron-scale metal powders such as (cold/hot) isostatic pressing, solid state sintering, chemically induced binding or liquid phase sintering of metal powders~\cite{Kruth2005}.\\

The remainder of this article is structured as follows: Section~\ref{sec:model} presents the main constituents of the proposed model, both, from a physical but also from a numerical point of view, with a special focus on cohesive effects. In Section~\ref{sec:model_calibration}, the proposed model is calibrated, i.e. the effective surface energy underlying the model is determined for the considered powder material by fitting numerical and experimental AOR values. In this section, also the sensitivity of the resulting AOR values, and the fitted surface energy, with respect to other powder material parameters and the spatial problem scaling is analyzed. Finally, a summary of the main results and a brief outlook on potential future research work is given in Section~\ref{sec:conclusion}.\\

\section{Physical and computational model}
\label{sec:model}

\begin{figure}[h!!]
 \centering
   {
    \includegraphics[height=0.33\textwidth]{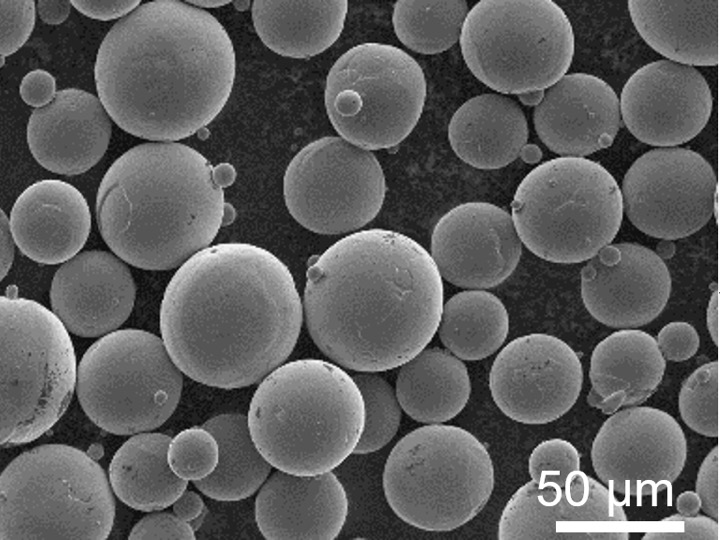}
    \label{fig:powder_SEM1}
   }
\hspace{0.5cm}
   {
    \includegraphics[height=0.33\textwidth]{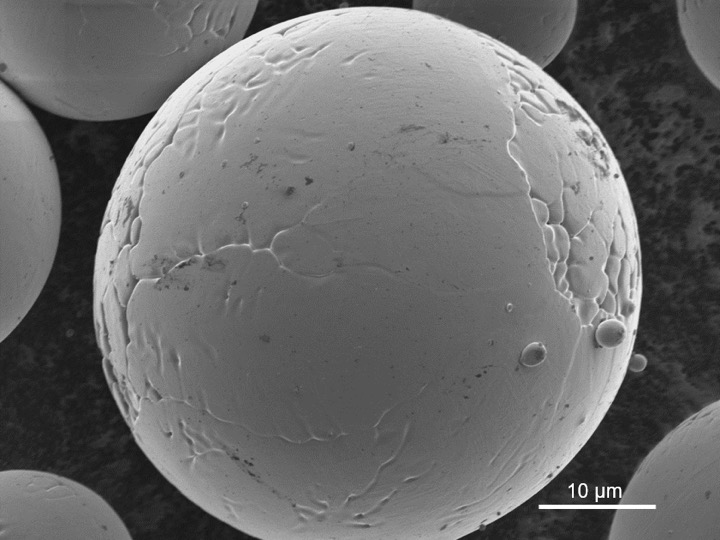}
    \label{fig:powder_SEM2}
   }
  \caption{SEM Images of employed plasma-atomized Ti-6Al-4V powder as typical for metal additive manufacturing processes.}
  \label{fig:powder_SEM}
\end{figure}

The experimental and numerical studies presented throughout this work employ plasma-atomized Ti-6Al-4V powder with different size distributions. Exemplary SEM images at different resolutions are given in Figure~\ref{fig:powder_SEM}. Accordingly, the shape of individual powder particles can be described as spherical in good approximation, an observation that could be confirmed by optical profilometer measurements conducted prior to this study. In the following, the discrete element method (DEM, originally developed by Cundall and Strack~\cite{Cundall1979}, see also~\cite{Bashira1991,Bolintineanu2014,Thornton2000}), more precisely the so-called "soft sphere" model, will be applied in order to model the bulk powder on the level of individual powder particles in a Lagrangian manner. This model describes powder particles as rigid spheres whose kinematics in 3D space are uniquely defined by six degrees of freedom, i.e. the position vector $\mb{r}_G$ as well as the rotation vector $\boldsymbol{\psi}$ representing the centroid position and orientation of the sphere, respectively. Accordingly, the equations of motion of an individual particle $i$ are given by the balance of linear and angular momentum:
\begin{subequations}
\label{momentum}
\begin{align}
(m \, \ddot{\mb{r}}_G)^i = m^i \mb{g} + \sum_j (\mb{f}_{CN}^{ij}+\mb{f}_{CT}^{ij}+\mb{f}_{AN}^{ij}),\label{gusarov2007_HCE1}\\
(I_G \, \dot{\boldsymbol{\omega}})^i = \sum_j (\mb{m}_{R}^{ij}+\mb{r}_{CG}^{ij} \times \mb{f}_{CT}^{ij}).\label{gusarov2007_HCE2}
\end{align}
\end{subequations}
Here, $m\!=\!4/3\pi r^3\rho$ is the particle mass, $I_G=0.4mr^2$ is the moment of inertia of mass with respect to the particle centroid $G$, $r$ is the particle radius, $\rho$ is the mass density, $\mb{g}$ is the gravitational acceleration, and $\dot{(...)}$ represents the first time derivative. For later use, also the effective mass and radius of two colliding particles $i$ and $j$ shall be defined as $m_{eff}:=\frac{m_i m_j}{m_i + m_j}$ and $r_{eff}:=\frac{r_i r_j}{r_i + r_j}$, respectively. All results derived in the following for the interaction of two spheres with masses $m_i,m_j$ and radii $r_i,r_j$ can be particularized for the interaction of two equal spheres ($m_i=m_j=m, \, r_i=r_j=r$) leading to  $m_{eff}=m/2, \, r_{eff}=r/2$ as well as for the interaction of a sphere with a rigid wall ($m_i=m, \, r_i=r, \, m_j \rightarrow \infty, \, r_j \rightarrow \infty$) leading to $m_{eff}=m, \, r_{eff}=r$. As consequence of the non-additivity of large rotations, the angular velocity $\boldsymbol{\omega}$ is in general different from the time derivative of the rotation vector $\boldsymbol{\psi}$, i.e. $\boldsymbol{\omega} \neq \dot{\boldsymbol{\psi}}$~\cite{Cardona1988,Meier2016,Simo1986}. However, due to the spherical shape of the particles and the type of interaction forces on the right-hand side of~\eqref{momentum}, the knowledge of the evolution of $\boldsymbol{\psi}(t)$ (which would require a special class of time integration schemes, so-called Lie-Group time integrators~\cite{Bruels2010,Romero2008b}) is not needed throughout the simulation. Instead, it is sufficient to solve for $\mb{r}_G$ and $\boldsymbol{\omega}$ as well as their time derivatives. In the present work, a velocity verlet time integration scheme is applied for this purpose. The right-hand side of~\eqref{momentum} summarizes the forces and torques resulting from an interaction with neighboring particles $j$ considering normal and tangential contact forces $\mb{f}_{CN}^{ij}$ and $\mb{f}_{CT}^{ij}$, adhesive forces $\mb{f}_{AN}^{ij}$, rolling resistance torques $\mb{m}_{R}^{ij}$ as well as the torques resulting from the tangential contact forces. In this context, $\mb{r}_{CG}^{ij}:=\mb{r}_{C}^{ij}-\mb{r}_{G}^i$ represents the vector from the centroid of particle $i$ to the point of contact with particle $j$. The specific force laws of these interactions will be discussed in the following section.\\

Besides powder modeling in a discrete sense as discussed so far, i.e. individual powder particles and their mechanical behavior are explicitly resolved, it would also be possible to describe the powder as a homogenized continuum based on a proper bulk powder material model (see e.g.~\cite{Erhart2005}). However, for the intended class of applications, namely the powder re-coating process in metal AM, the size of individual powder particles is not sufficiently small compared to the length scales of the problem domain (e.g. the ratio of powder particle diameter to powder layer thickness is typically in the range of $0.2-1$), i.e. a separation of length scales as required for a continuum approach is not given. Moreover, the analysis of certain effects of interest, e.g. groove-type defects in the powder layer caused by jammed powder particles, requires the resolution of individual particles as incorporated by the present DEM-based approach.\\

\subsection{Contact forces}
\label{sec:model_contact}

Since the employed DEM scheme considers rigid spheres, the compliance of particles during collision has to be accounted for by means of a proper contact force law reflecting the underlying material properties. Here, the normal contact force law is based on a typical spring-dashpot model of the form
\begin{align}
\label{normalcontactlaw1}
     \mb{f}_{CN}^{ij}=\left\{\begin{array}{ll}
                                 \min{(0,k_N g_N + d_N \dot{g}_N)}\mb{n}, & g_N \leq 0 \\
                                 \mb{0}, & g_N > 0
                    \end{array}\right.
\end{align}
where $g_N:=||\mb{r}_{G}^j-\mb{r}_{G}^i||-(r_i + r_j)$ is the normal gap between the contact points and $\mb{n}=(\mb{r}_{G}^j-\mb{r}_{G}^i)/||\mb{r}_{G}^j-\mb{r}_{G}^i||$ is the unit normal vector~\cite{Luding2006, Luding2008}. The elastic / penalty constant $k_N$ as well as the damping constant $d_N$ are given by:
\begin{align}
\label{normalcontactlaw2}
 d_N=2|\ln{(c_{COR})}|\sqrt{\frac{k_N m_{eff}}{\ln{(c_{COR})}^2+\pi^2}}, \quad \quad \quad k_N \geq \max{\left( \frac{8\pi\rho V_{max}^2 r_{max}}{c_{g}^2}, \frac{4\pi \gamma}{c_{g}} \right)}.
\end{align}
The case distinction as well as the $min()$ function in~\eqref{normalcontactlaw1} ensure that mechanical contact forces act only for negative normal gaps (penetration) and are always negative (tension cut-off), i.e. only pressure forces appear. The first equation in~\eqref{normalcontactlaw2} represents an analytic relation between the damping constant $d_N$ of the linear spring-dashpot (LSD) model and the coefficient of restitution $c_{COR}$ (which holds exactly as long as no tension cut-off is considered)~\cite{Herrmann1998}. The second equation states the minimal value of the elastic constant $k_N$ in order to guarantee for maximal relative penetrations $c_{g}:=\max_{i,t}{(|g_N|/R)}$ below a prescribed threshold value, considering dynamic collisions of particles with maximal radius $r_{max}$ and maximal velocity $V_{max}:=\max_{i,t}{(\dot{\mb{r}}_G^i)}$ (first term) as well as the maximal static adhesive forces resulting from the surface energy $\gamma$ (second term, see also Section~\ref{sec:model_adhesion}). Here, $\max_{i,t}{(...)}$ represents the maximum across all particles and the considered time span.\\

There exists a variety of more elaborate normal contact force models (see e.g.~\cite{Herrmann1998,Kruggel-Emden2007,Walton1986}), where the elastic part follows e.g. a $g_N^{(3/2)}$ law based on Hertzian contact theory~\cite{Hertz1881}, allowing for a relation of the elastic constant $k_N$ with the Young's modulus $E$ and the Poisson's ratio $\nu$ of the particle material. However, in DEM simulations, it is standard practice to choose the elastic constant by orders of magnitude lower as predicted by the Young's modulus in order to end up with larger critical time step sizes (see below) and, in turn, with shorter computation times. Since lower elastic constants lead to higher contact times and lower peak forces during dynamic collisions, this strategy does not allow for an exact representation of contact force evolutions in highly dynamic scenarios. On the other hand, it does allow to represent macroscopic / integrated quantities such as momentum exchange between particles as well as bulk powder kinematics / flow in an accurate manner as long as the relative penetration $c_{g}$ is reasonably small~\cite{Herbold2015,Luding2006}. For quasi-static problems (which applies to the majority of examples in this work), where inertia effects are negligible compared to gravity and adhesive forces, even the prediction of contact forces is possible in good approximation. These considerations also justify the application of the simple linear force law~\eqref{normalcontactlaw1} instead of more elaborate models.\\

In a similar manner, the tangential / frictional contact interactions are given by a Coulomb force law of the form
\begin{align}
\label{tangentcontactlaw1}
     \mb{f}_{CT}^{ij}=\left\{\begin{array}{ll}
                                 \min{(\mu ||\mb{f}_{CN}^{ij}||,|| k_T \mb{g}_T + d_T \dot{\mb{g}}_T ||)}\mb{t}_T, & g_N \leq 0 \\
                                 \mb{0}, & g_N > 0
                    \end{array}\right.
\end{align}
with a maximal force amplitude of $\mu ||\mb{f}_{CN}^{ij}|$ limited by the friction coefficient $\mu$ in the slip case and a regularization based on a linear spring-dashpot combination in the stick case~\cite{Luding2006, Luding2008}. Here, the unit tangent vector is defined as $\mb{t}_T:=- (k_T \mb{g}_T + d_T \dot{\mb{g}}_T )/|| k_T \mb{g}_T + d_T \dot{\mb{g}}_T ||$ and the rate of tangential gap vector is given by $\dot{\mb{g}}_T:=(\mb{I}-\mb{n} \otimes \mb{n}^T) (\mb{v}_{G}^i-\mb{v}_{G}^j) + \boldsymbol{\omega}_i \times \mb{r}_{CG}^{ij} - \boldsymbol{\omega}_j \times \mb{r}_{CG}^{ji}$, with $\mb{r}_{CG}^{ij}:=\mb{r}_{C}^{ij}-\mb{r}_{G}^i=(r_i+g_N/2)\mb{n}$, $\mb{r}_{CG}^{ji}:=\mb{r}_{C}^{ji}-\mb{r}_{G}^j=-(r_j+g_N/2)\mb{n}$, and with $\mb{I}$ representing an identity tensor in $\Re^3$. In contrast to the normal gap $g_N$, the tangential gap vector $\mb{g}_T$ is not uniquely defined by the current geometric configuration but has to be determined numerically via integration of $\dot{\mb{g}}_T$. Here, a Backward Euler scheme is employed for time integration of $\mb{g}_T$, and a typical return mapping algorithm is applied in order to check the stick/slip condition in every discrete time step~\cite{Giannakopoulos1989,Laursen1993,Wriggers1990}. Furthermore, the constants of the tangential contact interaction are chosen as
\begin{align}
\label{tangentcontactlaw2}
 d_T=d_N, \quad \quad \quad k_T=\frac{1-\nu}{1-0.5\nu} k_N,
\end{align}
where $\nu$ is Poisson's ratio of the particle material. As shown in~\cite{DiMaio2004}, normal and tangential force laws should not be chosen independently. Instead, the ratio of normal to tangential stiffness is a material property of the contacting bodies. It has been shown for the linear spring dash-pot model that a choice according to~\eqref{tangentcontactlaw2} leads to optimal results with respect to macroscopic characteristics of the particle collision such as the change of linear and angular momentum~\cite{DiMaio2004,DiMaio2005,DiRenzo2004}, which are comparable to the results gained by more elaborate (and complex) approaches. Again, since the advantages of these more elaborate models in terms of more accurate contact force evolutions would only come into play for a choice of stiffness parameters consistent with the actual material properties and since such a choice would lead to undesirably small time step sizes, the simple force laws~\eqref{normalcontactlaw1} and~\eqref{tangentcontactlaw1} seem to be a reasonable choice.

\subsection{Rolling resistance}
\label{sec:model_rolling}

The Coulomb friction law in combination with perfectly spherical particles as considered so far does not account for any kind of energy dissipation related to rolling motion.
For the considered Ti-6Al-4V powder particles, possible sources of rolling resistance might include viscous dissipation or elastic hysteresis losses in the contact zone, plastic deformation or deviations from the ideal spherical shape (at least on the scale of surface roughness asperities)~\cite{Brilliantov1998,Herbold2015,Zhou1999}. According to~\cite{Walton2008}, the particle radius below which plastic deformation would be expected solely due to adhesive interaction of two particles can be approximated by the relation $r_{cY}\approx 10^7 \gamma / E$. Taking the Young's modulus of Titanium and the maximal surface energy value of $\gamma=0.4 mJ /m^2$ considered in this work (see Section~\ref{sec:model_calibration}) leads to a critical particle diameter in the range of $0.1 \mu m$, which is far below the considered range of particle sizes. Thus, plastic deformation, at least of particle bulk material, seems to be negligible in good approximation. Based on the SEM images in Figure~\ref{fig:powder_SEM}, it has been decided that also the deviations from the ideal spherical shape do not have to be considered in what follows. Due to the absence of these two sources of "static" / ''constant" rolling resistance (see e.g.~\cite{Luding2006}), only a velocity-proportional, viscous contribution to rolling resistance of the form
\begin{align}
\label{rollingresistance1}
     \mb{m}_{R}^{ij}=-\mb{m}_{R}^{ji}= d_R ||\mb{f}_{CN}^{ij}|| r_{eff} \Delta \boldsymbol{\omega}_{\perp} \quad \text{with} \quad  \Delta \boldsymbol{\omega}_{\perp}=(\mb{I}-\mb{n} \otimes \mb{n}^T)(\boldsymbol{\omega}_i-\boldsymbol{\omega}_j)
\end{align}
will be considered. This form of a rolling resistance torque proportional to $r_{eff} \Delta \boldsymbol{\omega}_{\perp}$ follows from a reformulation of the objective and momentum-conserving rolling resistance formulation proposed in~\cite{Luding2006}. One very appealing approach of determining the coefficient $d_R$ has been proposed by Brilliantov et al.~\cite{Brilliantov1998}, who investigated the rolling motion of a deformable sphere on a rigid surface and derived a first-principle analytical model relating the rolling resistance coefficient to particle material properties according to
\begin{align}
\label{rollingresistance2}
     d_R \approx \frac{(1-c_{COR})}{C_1V_{I}^{1/5}}\left(  \frac{E \sqrt{r_{eff}/2}}{(1-\nu^2)} \right)^{-1/5} \quad \text{with} \quad C_1=1.15344
\end{align}
instead of taking it as an additional unknown parameter. Among others, this approximation is based on a Taylor series expansion of the coefficient of restitution $c_{COR}$ as a function of impact velocity $V_{I}$~\cite{Schwager1998}. While the velocity-dependence of $c_{COR}$ is well-documented in the literature~\cite{Herbold2015,Marinack2013,Wu2009}, it is a common approach to employ normal contact models such as the LSD force law~\eqref{normalcontactlaw1} that lead to a constant coefficient of restitution, which can in turn be prescribed as model input parameter based on a properly chosen reference velocity $V_{ref}$~\cite{DiMaio2004,DiMaio2005}. To stay consistent with the derivations in~\cite{Schwager1998}, the same reference velocity will be chosen as impact velocity $V_{I}=V_{ref}$ in~\eqref{rollingresistance2} (see Section~\ref{sec:model_parameters}). This procedure seems to be reasonable to get a first-order approximation for the otherwise unknown parameter $d_R$.

\subsection{Adhesive forces}
\label{sec:model_adhesion}

There are several physical sources of adhesive inter-particle forces, the most important being capillary effects, electrostatic potentials, and van der Waals (vdW) interactions. The latter term commonly refers to the accumulated effects resulting from so-called Keesom forces (interaction of two permanent dipoles), Debeye forces (interaction of a permanent and an induced dipole) as well as London dispersion forces (interaction of polarizable but initially non-polar molecules). Since metal powder fusion processes such as SLM require the powder humidity level to be as low as possible~\cite{Das2003,Meier2018}, and sometimes the powder is even vacuum-dried before application~\cite{Li2016}, perfectly dry powder will be considered here without explicitly accounting for any capillary effects. Furthermore, it can be assumed that van der Waals forces dominate electrostatic forces by orders of magnitude for \textit{conductive} powders with particle sizes below $100 \mu m$ as considered here~\cite{Castellanos2005,Israelachvili2011,Visser1989,Walton2008,Yang2000}. The following paragraphs summarize the most important basics required for the subsequent modeling of vdW interactions, and are in strong analogy to the review article by Walton~\cite{Walton2008}. A more detailed treatment of this topic can e.g. be found in the works by Isrealachvili~\cite{Israelachvili2011}, Castellanos~\cite{Castellanos2005} and Parsegian~\cite{Parsegian1997}.\\

Let's assume that the attractive potential between two atoms or small molecules due to van der Waals interaction is given by $\Psi_A(\tilde{s})=-C_f/\tilde{s}^6$, where $\tilde{s}$ is the distance between the atoms and $C_f$ is a constant. For elementary geometries, the resulting overall potential can be derived by summation / integration of $\Psi_A(\tilde{s})$ over all atoms within the considered volume. For the resulting interaction potential per unit area between two planar surfaces / half spaces (one surface area is limited to unity to keep the integral bounded) of distance $\tilde{s}$, this procedure yields:
\begin{align}
\label{adhesion1}
 w_P(s)=-\frac{\pi C_f \rho_a^2}{12s^2}=-\frac{A}{12 \pi s^2} \quad [J/m^2],
\end{align}
with $\rho_a$ being the atom / molecule density and $A = \pi^2 C_f \rho_a^2$ the Hamaker constant of the material~\cite{Hamaker1937}. In order to estimate the theoretical value of $w_P(s)$ for two (ideally smooth) planes in contact, the distance $s$ between the planes is typically set to the theoretical molecule diameter $s_0 \approx 4 \mathring{A} = 0.4nm$, resulting in:
\begin{align}
\label{adhesion2}
 w_P(s_0)=-\frac{A}{12 \pi s_0^2}=:2\gamma \quad [J/m^2].
\end{align}
Equation~\eqref{adhesion2} states the theoretical definition of the surface energy per unit area $\gamma$ of the material. Here, the factor of $2$ reflects the convention that $\gamma$ represents the surface energy per newly created surface area $2 F$ when two contacting bodies with contact area size $F$ are separated. In a similar manner, the total interaction potential (this time \textit{not} per unit area) between two spheres with radii $r_i$ and $r_j$ and closest distance / gap $s$ can be determined via integration of $\Psi_A(\tilde{s})$ over the volume of both spheres, which yields $w_S(s)=-Ar_{eff}/(6s) \,\, [J]$ with $r_{eff}=r_ir_j/(r_i+r_j)$ (see above). Based on this result, the associated interaction force between the two spheres reads
\begin{align}
\label{adhesion3}
 F_S(s)=\frac{d}{ds}w_S(s) = \frac{A r_{eff}}{6s^2} \quad [N].
\end{align}
Evaluating the interaction force~\eqref{adhesion3} at the (theoretical) contact distance $s_0$ yields the so-called pull-off force $F_{S0}$:
\begin{subequations}
\label{adhesion4}
\begin{align}
F_{S0} =& \frac{A r_{eff}}{6s_0^2} \label{adhesion4a},\\
     =& -4\pi \gamma r_{eff} \label{adhesion4b},
\end{align}
\end{subequations}
where from~\eqref{adhesion4a} to~\eqref{adhesion4b}, the quotient $A/s_0^2$ has been expressed by means of the surface energy $\gamma$ according to~\eqref{adhesion2}.\\

Several existing modeling approaches for adhesive interactions between spheres apply the force law~\eqref{adhesion3} limited by the constant cut-off value~\eqref{adhesion4a} for small distances $s<s_0$. On the one hand, the implementation of this strategy is rather straightforward since the two required constants $A$ and $s_0$ are tabulated for many materials. On the other hand, this approach has also a range of drawbacks, especially when considering fine metal powders. First, and most important, real surfaces are not ideally smooth. Surface asperities on rough surfaces keep the effective distance between closest molecule pairs at a much higher level than predicted by the theoretical molecule diameter $s_0$. As discussed by Walton~\cite{Walton2008}, van der Waals forces are of extreme short range nature, decreasing by two (four) orders of magnitude when increasing the (effective) particle distance from the theoretical value of $0.4nm$ to  $4nm$ ($40nm$), giving rise not only to drastically reduced magnitudes but also to considerably increased uncertainty of experimental pull-off forces and surface energies compared to the theoretical values~\eqref{adhesion4a} and ~\eqref{adhesion2}. Second, real surfaces are typically contaminated e.g. by oxidation layers resulting in lower effective Hamaker constants and due to~\eqref{adhesion4a} also in lower pull-off force values. Third, the attractive forces between \textit{metallic} particles cannot only be accounted for by the vdW potential but also by extremely short-range non-additive electron exchange interactions, also denoted as metallic bonds, relevant at separations below $0.5 nm$. These additional forces might lead to higher theoretical surface energy values for metals as predicted by~\eqref{adhesion2}. The short-range nature is even more pronounced than for vdW forces and makes this type of interactions extremely dependent on surface quality / roughness and even mutual metal lattice orientation~\cite{Ferrante1985,Israelachvili2011}. It can be concluded that experimentally measured values of pull-off forces and surface energies show strong variations and are typically by orders of magnitudes lower than the theoretical values~\eqref{adhesion4a} and~\eqref{adhesion2}.\\

For the reasons mentioned above, it is more practicable to either experimentally measure $\gamma$ or $F_{S0}$ (related by~\eqref{adhesion4b}) in a direct manner, e.g. via an atomic force microscope (AFM), or to determine these parameters by fitting experimental and numerical results based on properly chosen bulk powder experiments and metrics (see Section~\ref{sec:model_calibration}). Within this work, an adhesive force of the form $\mb{f}_{AN}^{ij}=F_S(s)\mb{n}$ based on the following force law will be applied:
\begin{align}
\label{adhesion5}
     F_S(s)=\left\{\begin{array}{ll}
      								F_{S0}=-4\pi \gamma r_{eff}, & g_N \leq  g_0\\
                                 \frac{A r_{eff}}{6s^2} , & g_0 < g_N < g^*\\
                                 0, & g_N \geq  g^*
                    \end{array}\right. 
                    \quad \text{with} \,\,
   g_0 := \sqrt{\frac{A r_{eff}}{6 F_{S0}}}, \quad g^* := \sqrt{\frac{1}{c_{FS0}}\frac{A r_{eff}}{6 F_{S0}}} = \frac{g_0}{\sqrt{c_{FS0}}}.
\end{align}
The model is uniquely defined by the surface energy $\gamma$ (measured experimentally) as well as the Hamaker constant (a typical tabulated value from the literature of $A=40\cdot 10^{-20} J$~\cite{Israelachvili2011} is taken here). Here, $g_0$ represents the distance at which the vdW force curve~\eqref{adhesion3} equals the pull-off force $F_{S0}$, and $g^*$ represents a cut-off radius at which the vdW has a relative decline of $c_{FS0}:=F_S(g^*)/F_{S0}$ with respect to $F_{S0}$ (taken as $c_{FS0}=1\%$ within this work). From a comparison of experimental and numerical angle of repose measurements (see Section~\ref{sec:model_calibration}), we deduce typical surface energy values in the range of $\gamma=0.1mJ/m^2$. From these exemplary parameter values and a typical mean particle diameter of $d=2r=34 \mu m$ as considered in the following, the remaining parameters in~\eqref{adhesion5} take on values of $F_{S0}\approx 10^{-8}N$, $g_0 \approx 0.2 \mu m$ and $g^* \approx 2 \mu m$. For comparison, the gravity force acting on a $34$ micron Ti-6Al-4V particle is $F_G \approx 10^{-9} N$, which means already one order of magnitude lower than the corresponding pull-off force.\\

As discussed in~\cite{Walton2008}, due to the short-range nature of vdW forces, these are expected to mainly affect existing contacts ($F_{S0}$ as resistance to contact break-up), but to have only small influence on the question if close particles will eventually come into contact or not. This statement underlines again the importance of reliable pull-off force values determined for the specific powder at hand, while deviations in the value taken for the Hamaker constant are rather negligible. In fact, based on additional simulations conducted in the context of this work, it has been concluded that the existence of a steady force transition interval $g_0<g_N<g^*$ leads to a numerically better posed problem instead of having a force jump from $0$ to $F_{S0}$, but that the specific functional representation of this transition (e.g. linear transition instead of vdW law) has only a minor effect on the resulting bulk powder behavior. On the other hand, it turned out that changing the magnitude of the surface energy $\gamma$, and thus of $F_{S0}$, say by a factor of two can lead to considerably different bulk powder behavior (see Section~\ref{sec:model_calibration}). This argumentation leads to a further advantage of directly measuring / calibrating $F_{S0}$ instead of taking theoretical values of $\gamma$ in~\eqref{adhesion4b} or of $A$ and $s_0$ in~\eqref{adhesion4a}: Even though the force-distance law of neglected effects such as humidity or electrostatic interaction (which might possibly be of a more long-range nature than vdW forces) is not considered in the present model, the influence of these effects on the pull-off force is implicitly taken into account by performing the experimental model calibration at the same environmental conditions as prevalent in the considered application, e.g. the powder recoating process in AM.\\

\hspace{0.5 cm}
\begin{minipage}{15.0 cm}
\textbf{Remark:} The relation between pull-off force and surface energy in~\eqref{adhesion4b} results from the so-called DMT (Derjaguin-Muller-Toporov) model considering the integrated interaction potential over two ideal, undeformed spheres. An alternative approach, the so-called JKR (Johnson-Kendall-Roberts) model derives the pull-off force under consideration of elastic deformation due to contact interaction and adhesive forces on the basis of the Hertzian theory and results in $F_{S0} = -3\pi \gamma r_{eff}$, which is only by $25\%$ lower than~\eqref{adhesion4b}~\cite{Walton2008}. Again, as long as $F_{S0}$ is determined experimentally via direct measurement / calibration,  the two approaches solely differ in the deduced surface energy, which is only of theoretical value but does not change the final force laws of the model (more precisely, the higher surface energy value deduced by the JKR model from a measured pull-off force, would again cancel out when inserting this surface energy into $F_{S0} = -3\pi \gamma r_{eff}$ to calculate the pull-off force for a specific particle size combination during the simulation).
\end{minipage}

\subsection{Simulation strategy and choice of parameters}
\label{sec:model_parameters}

All simulation results presented in this work rely on an implementation of the proposed DEM formulation in the in-house code BACI, a parallel multiphysics research code with finite element, particle and mesh-free functionalities, developed at the Institute for Computational Mechanics of the Technical University of Munich~\cite{Wall2018}. The set of ordinary differential equations~\eqref{momentum} is discretized in time by an explicit velocity verlet scheme. The time step size $\Delta t$ has been chosen according to the following critical step size estimate (see e.g.~\cite{OSullivan2004}):
\begin{align}
\label{stepsize}
     \Delta t \leq \Delta t_{crit}= 0.2 \sqrt{\frac{m_{min}}{k_N}},
\end{align}
with $m_{min}:=\min_{i}(m)$ representing the minimal mass across all particles. While a finite element discretization has been chosen to represent rigid walls, the modelled particle size distribution is considered to be of log-normal type with parameters fitted to the material certificate of the powder employed in the experiments. Specifically, the particle diameter distribution of the reference powder considered throughout this work is specified by $10\%$, $50\%$ and $90\%$ percentiles according to  $D10=20 \mu m$,  $D50=34 \mu m$ and $D90=44 \mu m$, representing a typical medium-sized powder for AM applications. After fitting the employed log-normal distribution to these specifications, the range of particle sizes considered in the model has been limited to lie within $D10$ and $D90$, i.e. very small particles below $d_{min}=20 \mu m$ as well as very large particles above $d_{max}=44 \mu m$ have not been considered for reasons of computational efficiency. The density of the Ti-6Al-4V particles is $\rho=4430 kg/m^2$, and the Hamaker constant has been taken as $A=40\cdot 10^{-20} J$~\cite{Israelachvili2011}.\\

The friction coefficient and the coefficient of restitution for particle-to-particle and particle-to-wall interaction have been chosen as $\mu=0.4$ and $c_{COR}=0.4$ in the present study. Further, the reference velocity in~\eqref{rollingresistance2} has been chosen in the range of the maximal impact velocities between powder particles expected for the funnel simulations in Section~\ref{sec:model_calibration}. Eventually, the penalty parameter $k_N$ has been chosen according to~\eqref{normalcontactlaw2}. For the range of parameters considered in this work, the adhesion-related term in~\eqref{normalcontactlaw2} based on a surface energy of $\gamma=0.1mJ/m^2$ and the choice $c_g=2.5\%$ resulted in a worst-case estimate of $k_N = 0.05N/m$. Since only the surface energy value will be calibrated on the basis of bulk powder experiments, additional verification simulations based on varied parameter values will be conducted in Section~\ref{sec:model_calibration} in order to assess the sensitivity of the resulting AOR values and the fitted surface energy with respect to the aforementioned parameter choice. It will turn out that the variation of these parameters (within a reasonable range) will typically have considerably lower influence on the bulk powder behavior than variations of the surface energy.\\

In the next section, the effective surface energy $\gamma_0$ required to calculate the pull-off force according to~\eqref{adhesion4b} will be determined for the considered Ti-6Al-4V powder by fitting experimental and numerical AOR values. In order to analyse the influence of cohesion on the bulk powder behavior and the resulting AOR values, additional funnel simulations will be carried out for the parameter choices $\gamma=4\gamma_0$ and $\gamma=0.25\gamma_0$. As shown in~\cite{Yang2000}, the characteristics (e.g. packing fraction, coordination number, surface uniformity) a layer of cohesive powder takes on in static equilibrium, can in good approximation be formulated as function of the dimensionless ratio of adhesive (pull-off) and gravity force, which scales quadratically with particle size according to $F_{\gamma}/F_G=(2\pi r_{eff}\gamma)/(mg) \sim \gamma / (\rho g r^2) $. Thus, the increase / decrease of surface energy $\gamma$ by a factor of $4$ is equivalent to the decrease / increase of particle size $r$ by a factor of $2$, which allows to evaluate the influence of powder particle size on the mechanical / kinematic bulk powder behavior.\\

\section{Model calibration}
\label{sec:model_calibration}

In order to calibrate the surface energy of the proposed DEM model, the angle of repose (AOR) has been determined for powder piles resulting from experimental and numerical funnel tests. To measure the AOR experimentally, powder was dispensed through a funnel with a discharge opening of $1mm$ diameter and a cone angle of $60^\circ$ onto the center of a cube with side length $a\!=\!a_0\!=\!10mm$ placed below the opening of the funnel. The (fixed) distance between the center of the cube and the opening of the funnel has been chosen as small as possible for each powder type to limit the powder impact velocity. The test was concluded when the powder pile covered the entire surface of the cube such that further addition of powder would result in no additional accumulation of powder. While the angle of repose is a well-defined metric of noncohesive granular materials, the angle of repose measured for highly cohesive powders might depend on the deformation history as well as on the problem size, which influences e.g. its avalanching behavior~\cite{Castellanos2005}. To address the problem of history dependence, it has been ensured in experiments and simulations that the powder is initially in a "loose" state, i.e. that it is not pre-compressed. In order to study the sensitivity with respect to problem size, additional experiments with down-scaled variants of the problem setup (cube with side lengths $a\!=\!a_0/2\!=\!5mm$ and $a\!=\!a_0/4\!=\!2.5mm$ at unchanged discharge opening) have been conducted. Moreover, three powders with different particle size distributions (PSD) have been employed in the experimental funnel studies discussed in the following. These are a medium-sized powder typical for SLM applications ($D10=20\mu m, D50=34\mu m, D90=44\mu m$, mean diameter $\bar{d}=34\mu m$), a rather coarse-grained powder ($D10=46\mu m, D50=72\mu m, D90=106\mu m$, mean diameter $\bar{d}=72\mu m$) as well as a rather fine-grained powder ($D10=8\mu m, D50=14\mu m, D90=23\mu m$, mean diameter $\bar{d}=14\mu m$).\\

\begin{figure}[h!!]
 \centering
   {
    \includegraphics[height=0.6\textwidth]{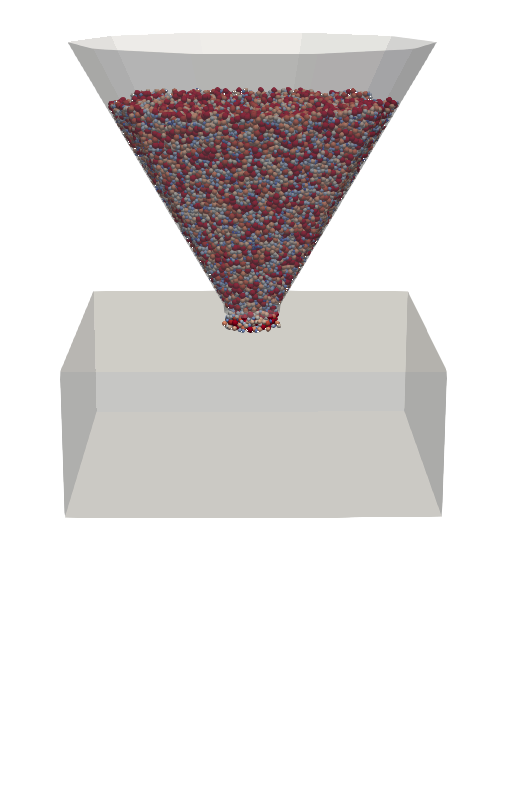}
    \label{fig:hallmodel_init}
   }
\hspace{1.5cm}
   {
    \includegraphics[height=0.6\textwidth]{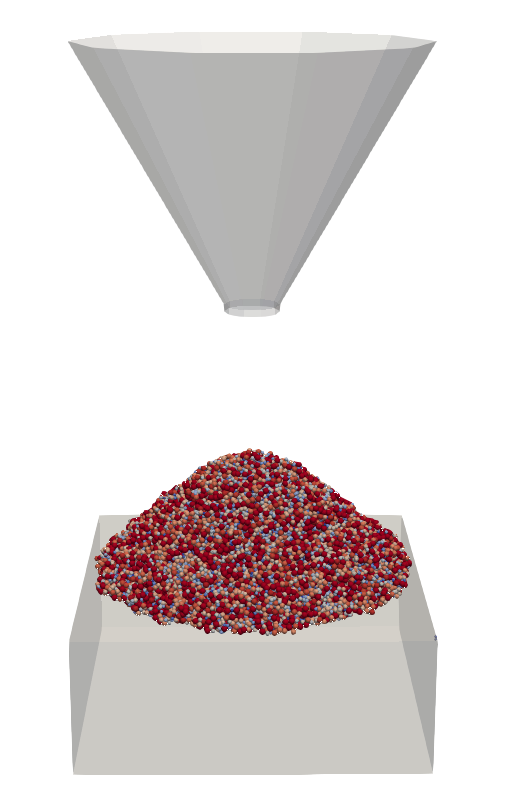}
    \label{fig:hallmodel_final}
   }
  \caption{Problem set up for numerical funnel tests in initial (left) and final (right) configuration.}
  \label{fig:hallmodel}
\end{figure}

In the numerical tests, the problem scaling $a\!=\!a_0/4$ has been considered to limit the computational costs. Furthermore, all numerical simulations within this work are based on a powder particle size distribution fitting the medium-sized powder ($\bar{d}=34\mu m$). In order to enable accurate AOR measurements also for the down-scaled numerical variant, it is desirable to keep the curved region at the particle impact zone on the top of the powder pile as small as possible, which can be achieved by reducing the powder mass flow and impact velocity, and thus the resulting impact momentum. In the numerical simulations, this is realized by driving down the cube during the flow experiments to limit the drop height of the particles. Furthermore, the diameter of the discharge opening has been decreased to $0.4mm$, and the adhesion between powder particles and the funnel walls has been switched off (while adhesion between particles and the cube is an important system property and has been kept) in order to ensure continuous powder flow even for this smaller opening. For the experimental and numerical analysis of very fine / cohesive powders, an additional small vibration amplitude has been applied to the funnel in order to increase flowability. It has been verified that all these modifications mainly influence the shape and relative size of the curved impact zone on the top of the powder pile without having noticeable influence on the angle of repose measured on the flanks of the pile. The initial and final configuration of the numerical problem setup is illustrated in Figure~\ref{fig:hallmodel}.\\

\begin{figure}[h!!]
   \centering
   \subfigure[$\gamma=0.0 \, mJ/m^2, \,\,\,\, AOR=11^\circ$.]
   {
    \includegraphics[width=0.3\textwidth]{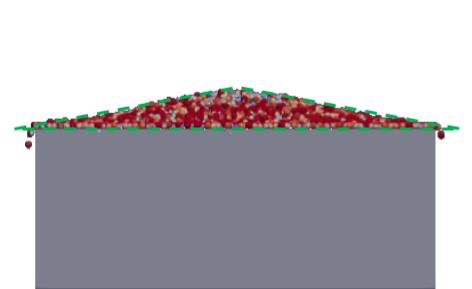}
    \label{fig:funnel_sim_0}
   }
   \centering
   \subfigure[$\gamma=0.01 \, mJ/m^2, \,\,\,\, AOR=24^\circ$.]
   {
    \includegraphics[width=0.3\textwidth]{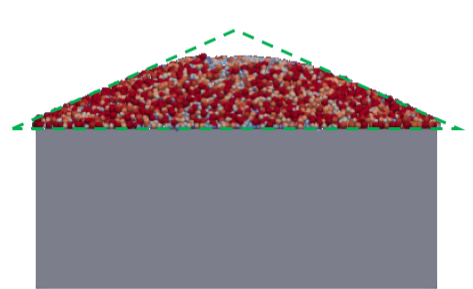}
    \label{fig:funnel_sim_1e8}
   }
    \centering
    \subfigure[$\gamma=0.02 \, mJ/m^2, \,\,\,\, AOR=29^\circ$.]
   {
    \includegraphics[width=0.3\textwidth]{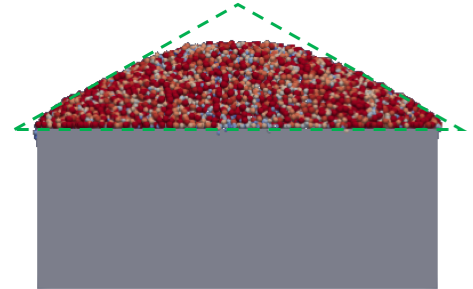}
    \label{fig:funnel_sim_2e8}
   }
    \centering
    \subfigure[$\gamma=0.04 \, mJ/m^2, \,\,\,\, AOR=33^\circ$.]
   {
    \includegraphics[width=0.3\textwidth]{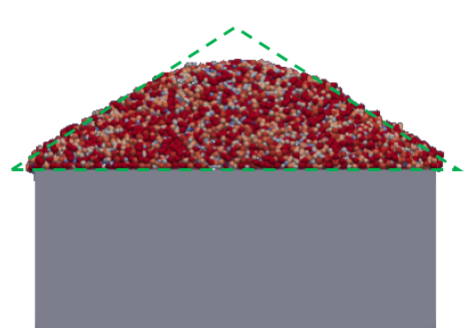}
    \label{fig:funnel_sim_4e8}
   }
   \centering
   \subfigure[$\gamma=0.06 \, mJ/m^2, \,\,\,\, AOR=34^\circ$.]
   {
    \includegraphics[width=0.3\textwidth]{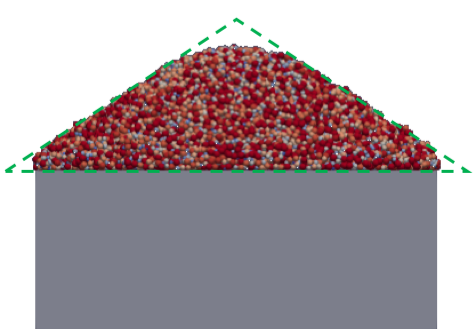}
    \label{fig:funnel_sim_6e8}
   }
    \centering
    \subfigure[$\gamma=0.08 \, mJ/m^2, \,\,\,\, AOR=37^\circ$.]
   {
    \includegraphics[width=0.3\textwidth]{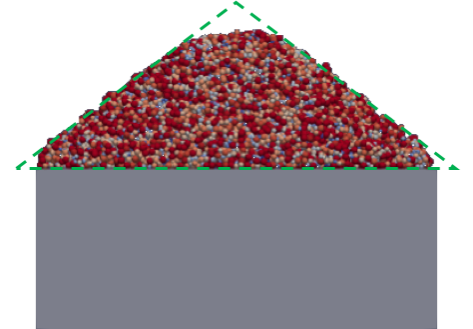}
    \label{fig:funnel_sim_8e8}
   }
    \centering
    \subfigure[$\gamma=0.1 \, mJ/m^2, \,\,\,\, AOR=41^\circ$.]
   {
    \includegraphics[width=0.3\textwidth]{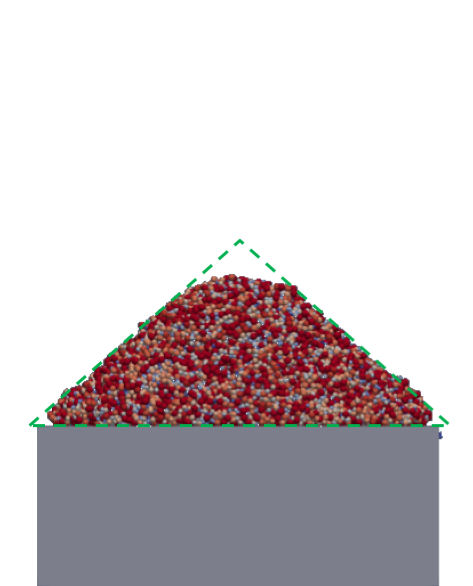}
    \label{fig:funnel_sim_1e7}
   }
 \centering
 \subfigure[$\gamma=0.2 \, mJ/m^2, \,\,\,\, AOR=57^\circ$.]
   {
    \includegraphics[width=0.3\textwidth]{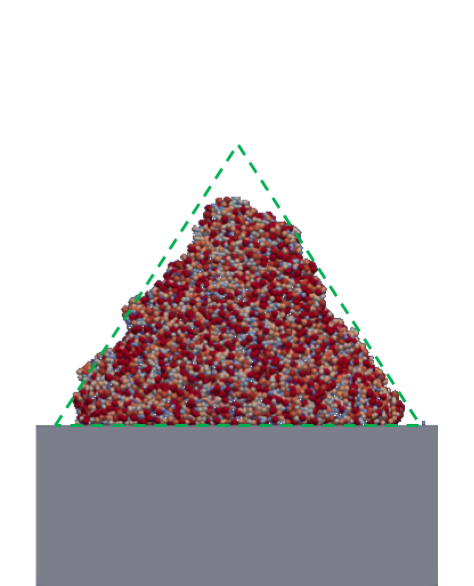}
    \label{fig:funnel_sim_2e7}
   }
   \centering
   \subfigure[$\gamma=0.4 \, mJ/m^2, \,\,\,\, AOR=63^\circ$.]
   {
    \includegraphics[width=0.3\textwidth]{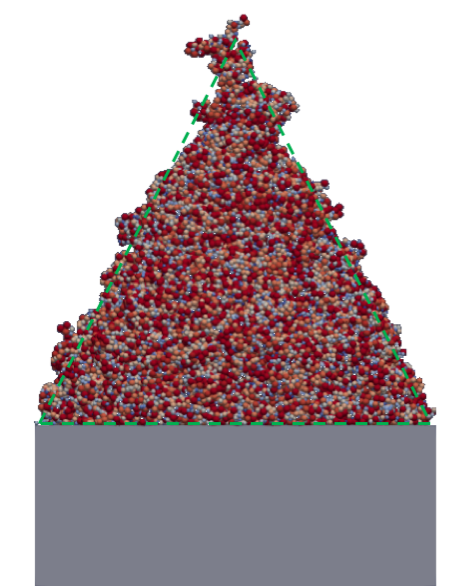}
    \label{fig:funnel_sim_4e7}
   }
   \caption{2D projections of the powder piles and derived AOR values resulting from numerical funnel simulations based on the medium-sized powder ($\bar{d}=34\mu m$) and different surface energies $\gamma$ .}
  \label{fig:funnel_sim}
\end{figure}

Figure~\ref{fig:funnel_sim} illustrates the 2D projections of the powder piles resulting from these numerical simulations for different choices of the surface energy $\gamma$. Based on these 2D images, the AOR values have been measured by fitting an equilateral triangle (green dashed lines) for each powder pile. As expected, the angle of repose increases with increasing surface energy, i.e. with increasing cohesiveness of the powder. Next, these numerical AOR values will be compared to the experimental counterparts measured for the medium-sized powder ($\bar{d}=34\mu m$) considered in the simulations.\\

\begin{figure}[h!!]
    \centering
    \subfigure[Side length cube: $a=a_0$ (16 samples).]
   {
    \includegraphics[height=0.21\textwidth]{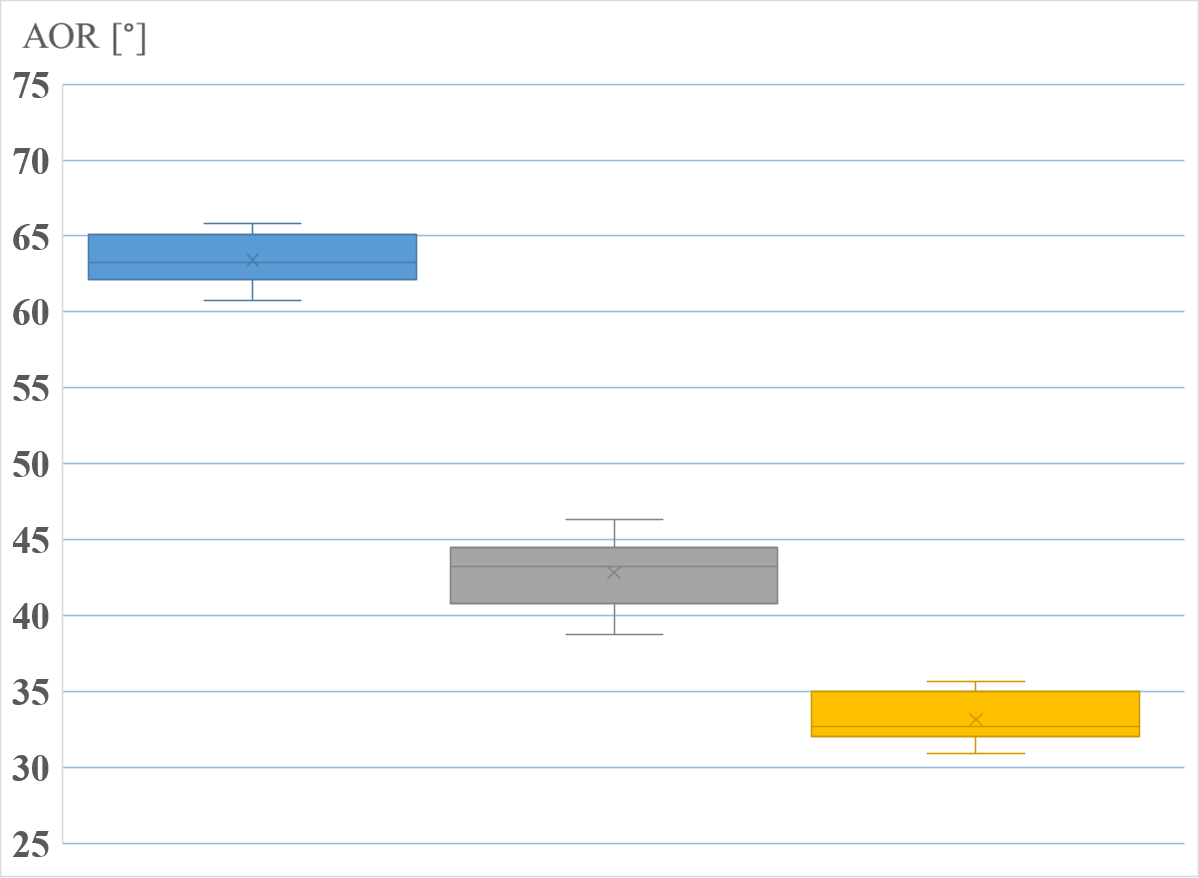}
    \label{fig:funnel_AOR_10mm}
   }
 \centering
 \subfigure[Side length cube: $a=a_0/2$ (3 samples).]
   {
    \includegraphics[height=0.21\textwidth]{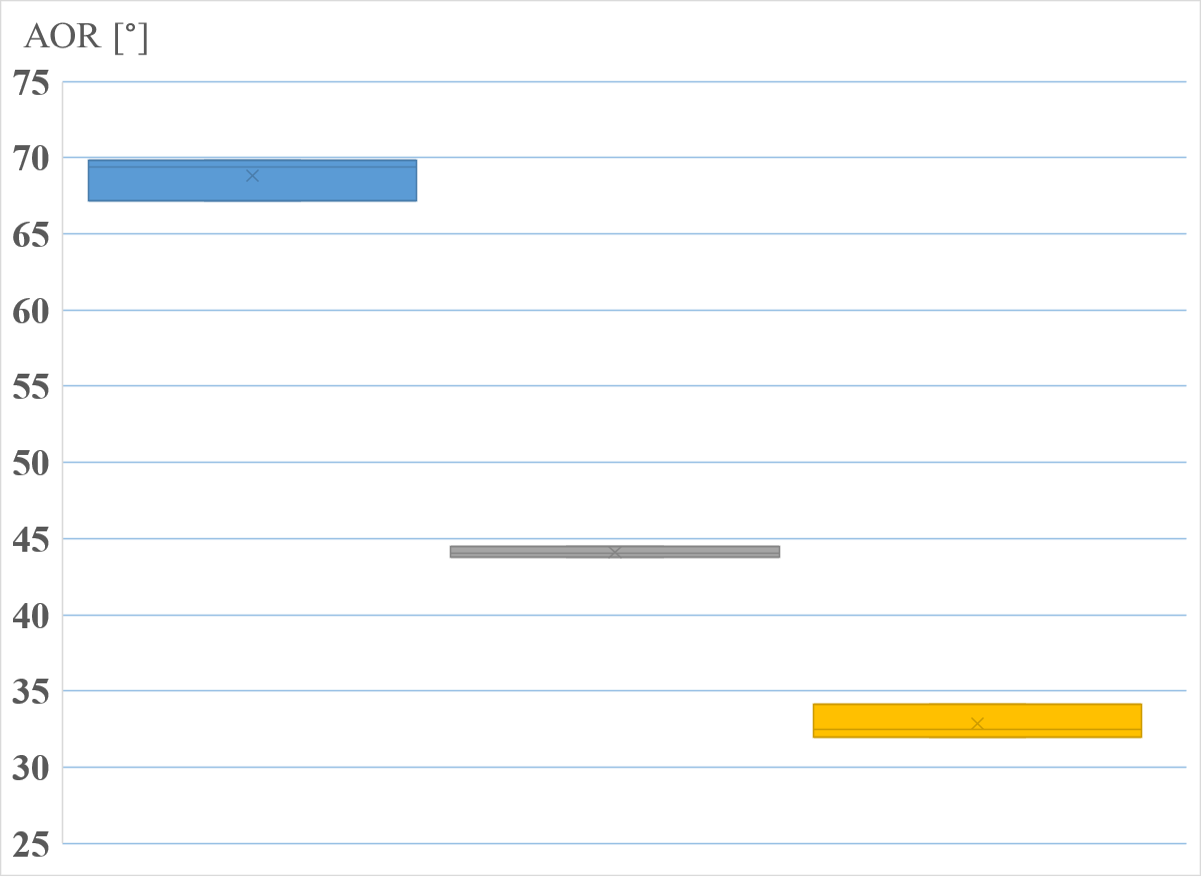}
    \label{fig:funnel_AOR_5mm}
   }
   \centering
   \subfigure[Side length cube: $a=a_0/4$ (3 samples).]
   {
    \includegraphics[height=0.21\textwidth]{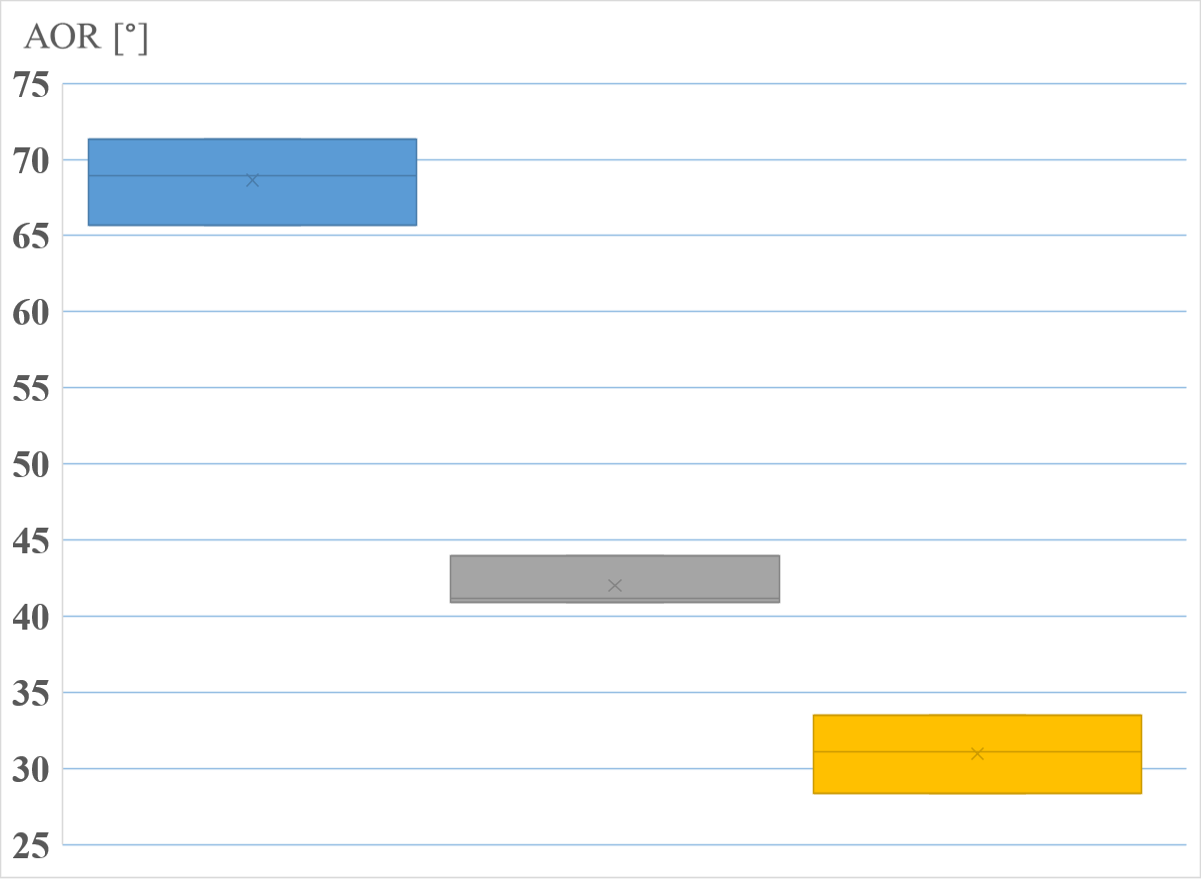}
    \label{fig:funnel_AOR_2_5mm}
   }
   {
    \includegraphics[height=0.21\textwidth]{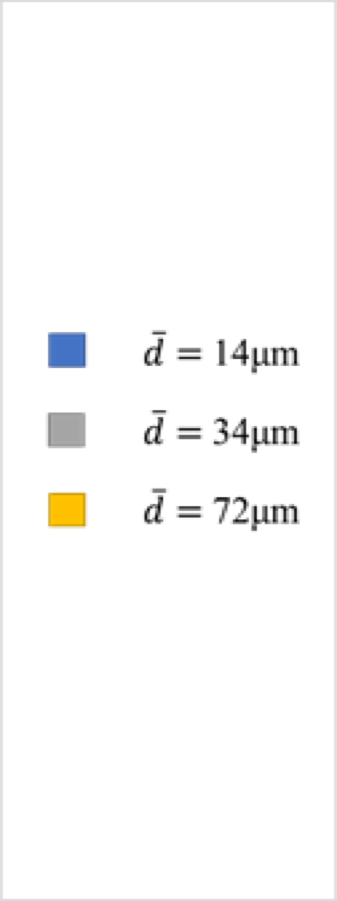}
   }
   \caption{Experimental AOR values for the three different powder size classes $\bar{d}$ and the three different problem scalings (cube side lengths $a$).}
  \label{fig:funnel_exp_statistics}
\end{figure}

For each of the three powder size classes ($\bar{d}=14\mu m, \,\, \bar{d}=34\mu m, \,\, \bar{d}=72\mu m$) as well as for each of the three problem scalings ($a\!=\!a_0, \,\, a\!=\!a_0/2, \,\, a\!=\!a_0/4$), several funnel experiments with different powder samples have been conducted. The measurement of the experimental AOR values is also based on 2D powder pile images and equivalent to the numerical strategy. A statistical evaluation of the measured AOR values is presented in Figure~\ref{fig:funnel_exp_statistics}. As expected, adhesion forces increasingly dominate gravity when decreasing the mean powder particle size, which results in higher AOR values. Additionally, it can be observed that the variation of the results for a given particle size distribution are in a reasonable range when comparing the different problem scalings or when comparing different powder samples for one problem scaling. For the medium and coarsed-grained powder, the variation of the mean AOR values lie wihin a range of approximately $3^{\circ}$ among the three different problem scalings. As expected, the most cohesive, fine-grained powder shows a stronger problem size-dependence leading to a variation of mean AOR values in the range of $7^{\circ}$ among the three problem scalings. However, since the medium-sized powder ($\bar{d}=34\mu m$), and not the fine-grained powder, will be employed for model calibration in the following, this observation has no negative effect on the accuracy of the fitted surface energy values. Eventually, it can be seen that also the variations among different powder samples of one problem and powder size are typically in a range below $3^{\circ}$. Only, the fine-grained as well as the coarse-grained powder in combination with the smallest problem scaling $a\!=\!a_0/4$ lead to larger variations in the range of $6^{\circ}$. This observation might be explained by the fact that the average particle size in the coarse-grained powder is rather large compared to the representative problem volume associated with the problem scaling $a\!=\!a_0/4$ (i.e. comparatively high Knudsen number $\bar{d}/a \approx 0.03$), which leads to stronger variations of results when analyzing effective, bulk powder properties on the continuum level. The same argumentation also holds for the fine-grained powder, but in this case not the size of individual particles but rather the size of particle agglomerates present in highly cohesive powders is relevant. Moreover, the high uncertainty and strong variation of effective adhesive forces in real powders (e.g. due to variations of surface roughness, contamination, mutual particle orientation) also contribute to the higher variation of AOR values observed for highly cohesive, small-grained powders. Again, the medium-sized powder, which is used for model calibration in the following, shows a variation below $3^{\circ}$ even for the smallest problem scaling $a\!=\!a_0/4$. Finally, for verification of the presented experimental procedure, it shall be stated that experimental AOR values presented by Sun et al.~\cite{Sun2015} for a comparable Ti-6Al-4V powder with highly spherical particles and a size range  that is identical to the investigated coarse-grained powder ($D10=46\mu m, D90=106\mu m$) led to average AOR values of $32^{\circ}$, which is in very good agreement with Figure~\ref{fig:funnel_exp_statistics}. Interestingly, Sun et al.~\cite{Sun2015} showed also that a Ti-6Al-4V powder with identical size range, but considerably reduced sphericallity (i.e. higher effective rolling resistance) led to an AOR value that is only by four degrees higher than for the aforementioned highly spherical Ti-6Al-4V particles. This observation suggests that cohesion is the dominating factor and has a higher influence on the resulting AOR values, and the bulk behavior of the considered Ti-6Al-4V powder in general, than the magnitude of and specific model for rolling resistance.\\

According to Figure~\ref{fig:funnel_exp_statistics}, the experimental results of the variant $\bar{d}=34\mu m$ and $a\!=\!a_0/4$, i.e. the variant that has also been considered in the numerical simulations, lead to a mean AOR of $\approx 41 ^{\circ}$. According to Figure~\ref{fig:funnel_sim}, this results fits very well to the simulation results based on a surface energy of $\gamma=0.1 \, mJ/m^2$, which thus will be considered as (experimentally fitted) estimate for the effective surface energy $\gamma_0=0.1 \, mJ/m^2$ of the powder from now on. While in the present study the mean experimental AOR value was already reasonably close to the AOR value of one of the numerical parameter variants (i.e. the case $\gamma=0.1 \, mJ/m^2$), an interpolation of the AOR values associated with the different numerical parameter variants will be required for this fitting strategy in general.\\

\begin{figure}[h!!]
    \centering
   \subfigure[Sim.: $\gamma\!=\!4\gamma_0$, $a\!=\!a_0/4$.]
   {
    \includegraphics[width=0.23\textwidth]{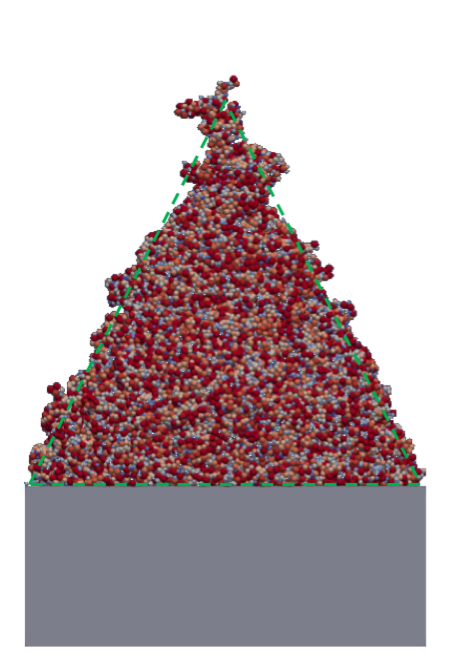}
    \label{fig:funnel_comp_fine_sim}
   }
 \centering
 \subfigure[Exp.: $\bar{d}\!=\!14\mu m$, $a\!=\!a_0/4$.]
   {
    \includegraphics[width=0.23\textwidth]{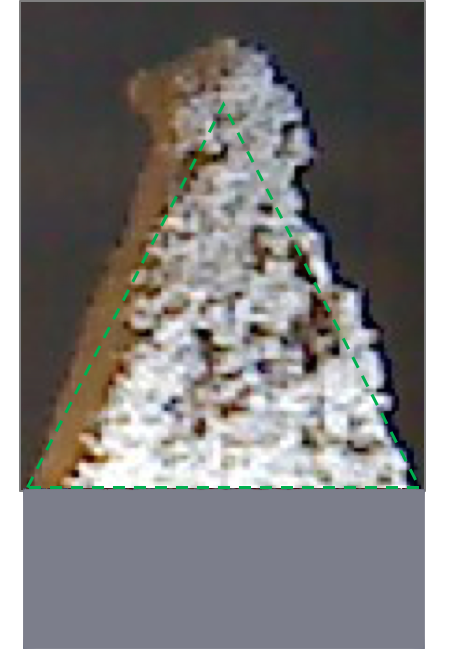}
    \label{fig:funnel_comp_fine_exp_25}
   }
   \centering
   \subfigure[Exp.: $\bar{d}\!=\!14\mu m$, $a\!=\!a_0/2$.]
   {
    \includegraphics[width=0.23\textwidth]{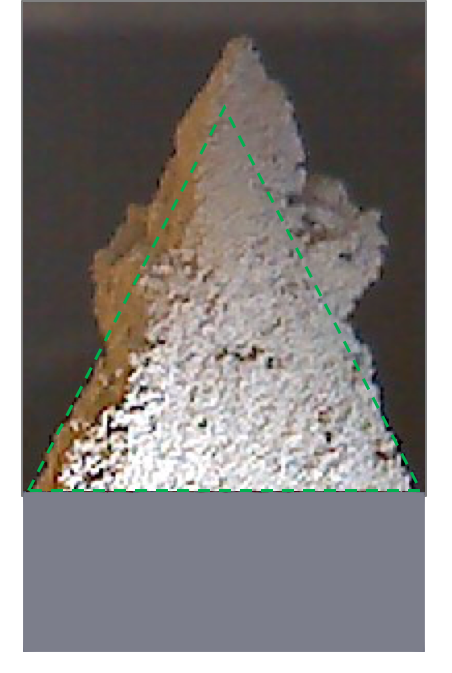}
    \label{fig:funnel_comp_fine_exp_50}
   }
    \centering
   \subfigure[Exp.: $\bar{d}\!=\!14\mu m$, $a\!=\!a_0$.]
   {
    \includegraphics[width=0.23\textwidth]{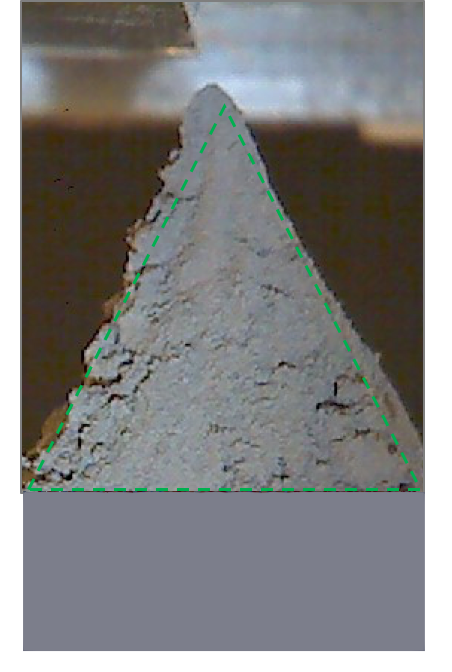}
    \label{fig:funnel_comp_fine_exp_100}
   }
    \centering
   \subfigure[Sim.: $\gamma\!=\!\gamma_0$, $a\!=\!a_0/4$.]
   {
    \includegraphics[width=0.23\textwidth]{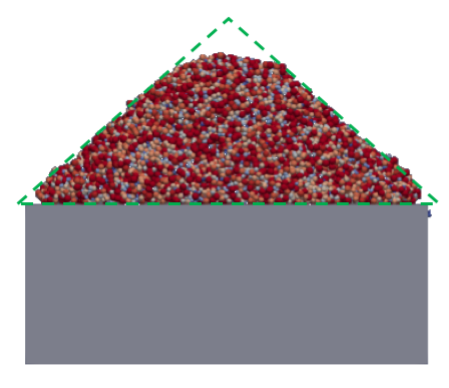}
    \label{fig:funnel_comp_medium_sim}
   }
 \centering
 \subfigure[Exp.: $\bar{d}\!=\!34\mu m$, $a\!=\!a_0/4$.]
   {
    \includegraphics[width=0.23\textwidth]{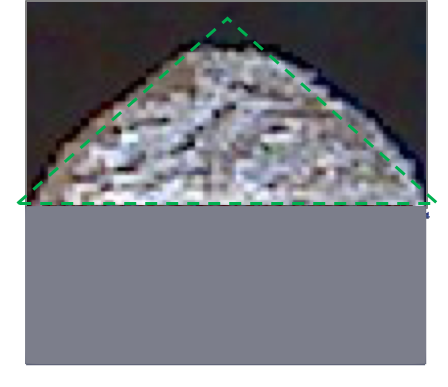}
    \label{fig:funnel_comp_medium_exp_25}
   }
   \centering
   \subfigure[Exp.: $\bar{d}\!=\!34\mu m$, $a\!=\!a_0/2$.]
   {
    \includegraphics[width=0.23\textwidth]{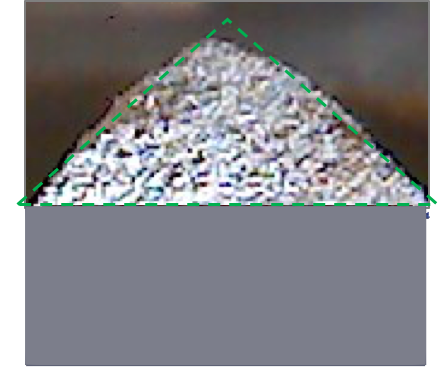}
    \label{fig:funnel_comp_medium_exp_50}
   }
    \centering
   \subfigure[Exp.: $\bar{d}\!=\!34\mu m$, $a\!=\!a_0$.]
   {
    \includegraphics[width=0.23\textwidth]{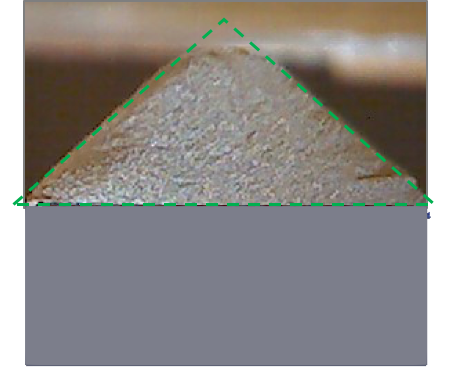}
    \label{fig:funnel_comp_medium_exp_100}
   }
    \centering
   \subfigure[Sim.: $\gamma\!=\!\gamma_0/4$, $a\!=\!a_0/4$.]
   {
    \includegraphics[width=0.23\textwidth]{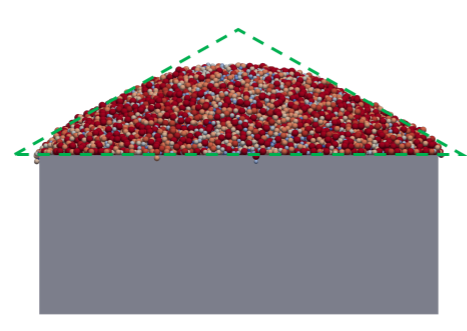}
    \label{fig:funnel_comp_rough_sim}
   }
 \centering
 \subfigure[Exp.: $\bar{d}\!=\!72\mu m$, $a\!=\!a_0/4$.]
   {
    \includegraphics[width=0.23\textwidth]{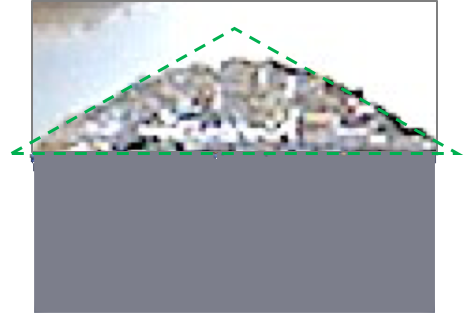}
    \label{fig:funnel_comp_rough_exp_25}
   }
   \centering
   \subfigure[Exp.: $\bar{d}\!=\!72\mu m$, $a\!=\!a_0/2$.]
   {
    \includegraphics[width=0.23\textwidth]{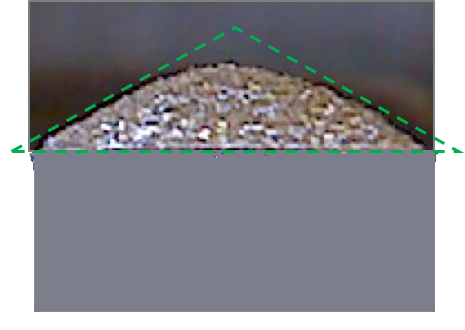}
    \label{fig:funnel_comp_rough_exp_50}
   }
    \centering
   \subfigure[Exp.: $\bar{d}\!=\!72\mu m$, $a\!=\!a_0$.]
   {
    \includegraphics[width=0.23\textwidth]{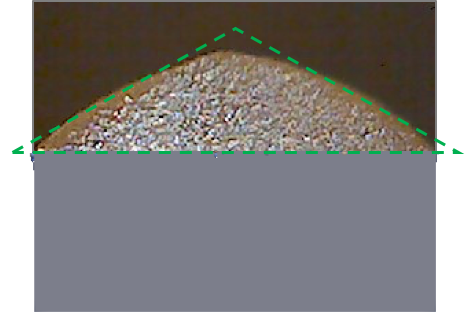}
    \label{fig:funnel_comp_rough_exp_100}
   }
   \caption{Numerical funnel results for powders with different surface energy $\gamma$ ($\gamma_0=0.1mJ/m^2$) as well as experimental funnel results for powders of different mean particle diameter $\bar{d}$ and different problem scalings (base cubes of different edge length $a$ with $a_0=10mm$).}
  \label{fig:funnel_sim_exp}
\end{figure}

As already discussed in the last section, the ratio of adhesion to gravity force increases quadratically with decreasing particle size. Thus, the effect of increasing / decreasing the mean particle diameter by a factor of two can be modeled by decreasing / increasing the surface energy by a factor of four at otherwise unchanged parameters. To verify this hypothesis, the parameter variants $\gamma\!=\!\gamma_0, \,\, \gamma\!=\!\gamma_0/4, \,\, \gamma\!=\!4 \gamma_0$ will be considered in the following (see first column of Figure~\ref{fig:funnel_sim_exp}). Experimentally, the equivalent cases of increasing / decreasing the mean particle diameter by a factor of two is approximately captured by comparing the coarse-grained (increasing by a factor $2.1$, third row of Figure~\ref{fig:funnel_sim_exp}) and fine-grained (decreasing by a factor $2.4$, first row of Figure~\ref{fig:funnel_sim_exp}) powder described above with the medium-sized powder (factor $1.0$, second row of Figure~\ref{fig:funnel_sim_exp}), which has been taken as reference for model calibration. The adhesion-to-gravity force ratios for the cases $\gamma\!=\!\gamma_0, \,\, \gamma\!=\!4\gamma_0, \,\, \gamma\!=\!\gamma_0/4, \,\, \gamma\!=\!0$ as well as for the corresponding equivalent mean particle diameters $\bar{d}=d_0, \,\, \bar{d}=d_0/2, \,\, \bar{d}=2d_0, \,\, \bar{d} \rightarrow \infty$, with $d_0=34\mu m$, are illustrated in Table~\ref{tab:equivalent_gamma_vs_radius}. From Table~\ref{tab:equivalent_gamma_vs_radius}, it becomes obvious that the cases $\gamma = \gamma_0$ and $\gamma = 4 \gamma_0$ are characterized by adhesion forces that are one to two orders of magnitude higher than the gravity forces. In this regime, the cohesiveness of the powder is already expected to have considerable influence on the bulk powder characteristics~\cite{Yang2000}.\\

\begin{table}[h!]
\centering
\begin{tabular}{|p{3.0cm}|p{1.5cm}|p{1.5cm}|p{1.5cm}|p{1.5cm}|} \hline
 $\bar{d}=d_0, \quad \quad \quad  \gamma / \gamma_0:$ & $0$ & $0.25$ & $1$ & $4$ \\ \hline
$\gamma = \gamma_0, \quad \quad \quad  \bar{d} / d_0:$ & $\infty$ & $2$ & $1$ & $0.5$ \\ \hline
$\quad \quad \quad  \quad \, \quad F_{\gamma}/F_G:$ & $0$ & $3.25$ & $13$ & $52$ \\ \hline
\end{tabular}
\caption{Surface energies and equivalent mean particle diameter and adhesion-to-gravity force ratios.}
\label{tab:equivalent_gamma_vs_radius}
\end{table}

Besides these different experimental and numerical parameter cases, Figure~\ref{fig:funnel_sim_exp} also visualizes the effect of the different experimental problem scalings $a\!=\!a_0/4, \,\, a\!=\!a_0/2, \,\, a\!=\!a_0$ (second to fourth column). For better comparison, the triangles fitted to the numerical results for AOR measurement are indicated in the experimental images as well. Experimental and numerical results show good qualitative agreement. Specifically, it can be observed that increasing adhesion / decreasing particle size leads to: increasing angle of repose and powder pile height; increasing surface roughness as well as tendency and size of cluster formation; decreasing size of curved impact zone at the top of the powder pile. The latter observation might be explained by the higher mass / momentum of larger particles when impacting on the powder pile and the lower adhesion forces (relative to gravity) attracting incident particles and impeding their recoil. From a quantitative point of view, the experimental AOR values of the fine-grained and coarse-grained powder can be compared with the equivalent numerical cases for verification purposes, while the case of the medium-sized powder has already been taken for model calibration. The numerical and experimental AOR values associated with the coarse-grained powder (third line in Figure~\ref{fig:funnel_sim_exp}) are in good agreement. This observation confirms the initial proposition that the effect of increasing the mean powder particle size by a certain factor is equivalent to decreasing the surface energy by the square of this factor. The experimental AOR values measured for the fine-grained powder, however, are slightly higher than their numerical counterparts. This observation can be explained by the fact that the mean particle diameter of the fine-grained powder is not exactly by a factor of $2$, but rather by a factor of $2.4$ smaller than the mean particle diameter of the medium-sized reference powder, which is equivalent to an increased surface energy by a factor of $2.4^2 \approx 5.8$. Furthermore, it becomes obvious again that due to the arguments stated above (strong variation of surface energies / adhesive forces, cohesive particle clusters large compared to size of powder pile) the AOR values as well as the powder pile morphology show stronger variations for the fine-grained powder.\\

It is also interesting to realize how drastically the AOR is underestimated in case no cohesion is considered at all (see Figure~\ref{fig:funnel_sim_0}), even in comparison with the coarse-grained powder variant (see third line in Figure~\ref{fig:funnel_sim_exp}). On the one hand, this might be explained by the employed velocity-proportional, viscous rolling resistance model~\eqref{rollingresistance1}, which has no contribution in static equilibrium configurations. There exist alternative, velocity-independent rolling resistance models (similar to sliding friction, see e.g.~\cite{Luding2006}), which can certainly have influence on the static AOR values of powder piles~\cite{Zhou1999,Zhou2002}. On the other hand, considering the highly spherical particle shapes of the plasma-atomized Ti-6Al-4V powder (see Figure~\ref{fig:powder_SEM}) and the amount of plastic surface deformation predicted for this powder (see also Section~\ref{sec:model_rolling}), it is expected that a physically reasonable choice of the rolling resistance coefficient underlying such a model still leads to a rather limited mechanical contribution. Thus, it can be concluded that the consideration / neglect of cohesion indeed makes the essential difference between realistic bulk powder behavior as observed in Figure~\ref{fig:funnel_comp_rough_sim} and rather unphysical behavior as observed in Figure~\ref{fig:funnel_sim_0}.\\

\begin{figure}[h!!]
    \centering
   \subfigure[Standard parameter set.]
   {
    \includegraphics[width=0.23\textwidth]{Figs/funnel_sim_medium.png}
    \label{fig:funnel_sim_paramvar_orig}
   }
 \centering
 \subfigure[Modified.: $k_N=2 k_{N0}$.]
   {
    \includegraphics[width=0.23\textwidth]{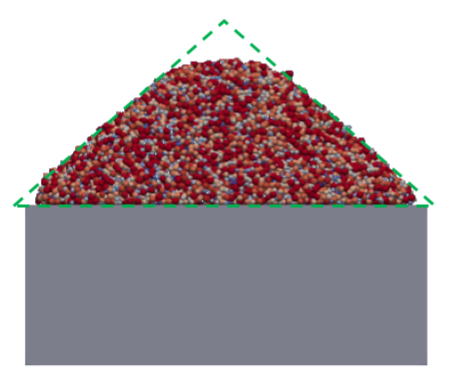}
    \label{fig:funnel_sim_paramvar_fric}
   }
   \centering
   \subfigure[Modified.: $\mu=1.5\mu_0$.]
   {
    \includegraphics[width=0.23\textwidth]{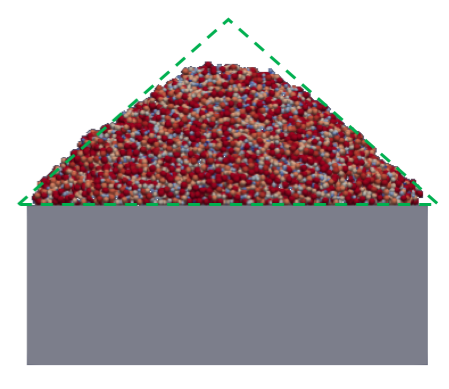}
    \label{fig:funnel_sim_paramvar_pen}
   }
    \centering
   \subfigure[Modified.: $c_{COR}=1.5c_{COR,0}$.]
   {
    \includegraphics[width=0.23\textwidth]{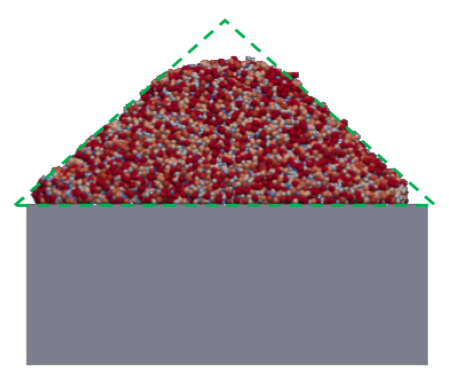}
    \label{fig:funnel_sim_paramvar_cor}
   }
   \caption{Influence of varying penalty parameter, friction coefficient and coefficient of restitution on resulting AOR ($\gamma_0=0.1mJ/m^2$).}
  \label{fig:funnel_sim_paramvar}
\end{figure}

Apart from rolling resistance, also the sensitivity concerning the choice of other model parameters has been investigated in the present study. According to Figure~\ref{fig:funnel_sim_paramvar}, the influence of increasing the penalty parameter by a factor of two, the friction coefficient by a factor of $1.5$ or the coefficient of restitution by a factor of $1.5$ is rather small compared to the influence of varied surface energy (see Figure~\ref{fig:funnel_sim}). The high sensitivity of the bulk powder behavior concerning the magnitude of surface energy as well as the strong variation of effective surface energy values for a given material pairing suggest to calibrate this model parameter on the basis of bulk powder experiments and simulations, while typical literature values can be taken for the remaining (less sensitive) material parameters in good approximation.\\

Finally, it shall be emphasized that the fitted value $\gamma_0=0.1mJ/m^2$ for the effective surface energy is considerably lower than typical values that can be found for metal surfaces in the literature. While the theoretical value for metals is in the range of $1000 mJ/m^2$, available experimental results measured for flat metallic contact surfaces are rather in the range of $30-50 mJ/m^2$~\cite{Faille2002,Guillemot2006,Hedberg2014,Herbold2015,Santos2004}, which is still two orders of magnitude higher than the fitted value $\gamma_0=0.1mJ/m^2$ found in the present study. The authors' ongoing research work focuses on verification of this result on the particle level via atomic force microscope (AFM) experiments. First AFM results suggest surface energies slightly above the fitted value $\gamma_0=0.1mJ/m^2$, but still in the same order of magnitude, i.e. below $1 mJ/m^2$.\\

Two conclusions can be drawn from this result: First, in order to improve the matching of single-particle (AFM experiments) and bulk powder (AOR measurements) predictions for the surface energy, the proposed adhesion force model might be further improved e.g. by accounting for the variation of the effective surface energy across different particle pairings or for a potential contact pressure-dependence of pull-off forces. However, already the present model (without these potential extensions) can be expected as sufficiently accurate for many applications since the model parameter "surface energy" has been fitted such that the experimentally observed bulk powder behavior could be matched. The fact that the parameter choice $\gamma_0=0.1mJ/m^2$ leading to an optimal representation of the experimentally observed bulk powder behavior depends on the specific adhesion force model (and, thus, might slightly deviate from corresponding AFM results), has been reflected in the present work by denoting $\gamma_0$ as \textit{estimate for the effective} surface energy. Second, in agreement with the AFM measurements performed by the authors so far, there is still strong evidence that the surface energy values prevalent for the considered class of metal powders are considerably lower than experimental values that can be found for flat metal contacts so far. As discussed in Section~\ref{sec:model_adhesion}, adhesive forces can easily vary by orders of magnitude depending on the geometrical and chemical properties of the contact surfaces (see also~\cite{Rabinovich2002}). The present results suggest that factors of influence such as surface roughness and potential chemical surface contamination / oxidation of the considered Ti-6Al-4V powder - which is known for its sensitivity concerning moisture / oxidation - reduce the surface energy to values that are rather in the range of $0.1-1 mJ/m^2$. These findings are expected to have considerable implications in applications based on fine metal powders e.g. concerning the powder layer uniformity in metal AM processes such as selective laser melting, electron beam melting or binder jetting~\cite{Meier2018b}. Based on the proposed calibration procedure, the powder model can easily be tailored to alternative powder materials.\\

\section{Conclusion}
\label{sec:conclusion}

The present work proposed a novel modeling and characterization approach for micron-scale metal powders relevant for a broad range of applications, with a special focus on powder-bed fusion additive manufacturing processes. The model is based on the discrete element method (DEM) and considers particle-to-particle and particle-to-wall interactions involving frictional contact, rolling resistance and adhesive forces. A special emphasis was on the modeling of cohesive effects, with an adhesion force law defined through the adhesive pull-off force in combination with a van-der-Waals force curve regularization. While most model parameters are not particularly sensitive and hence have been taken from the literature, the pull-off force defined by the effective surface energy between particles can easily vary by several orders of magnitude and has, to the best of the authors' knowledge, not been measured for the considered class of metal powder so far. In order to close this crucial gap, we have performed respective experiments and fitted angle of repose (AOR) values from numerical and experimental funnel tests for the considered, medium-sized Ti-6Al-4V powder (mean particle diameter $\bar{d}_0=34\mu m$), which allowed to estimate the associated effective surface energy in good approximation. Interestingly, the fitted surface energy value of $\gamma_0=0.1mJ/m^2$ is by four orders of magnitude lower than theoretical values and by two orders of magnitude lower than typical experimental values measured for flat metal contacts. Potential surface contamination / oxidation as well as increased surface roughness might be responsible for the decay of this extreme short-range interaction.\\

Besides the fitted surface energy value $\gamma=\gamma_0$, also the cases of decreased / increased surface energies $\gamma=\gamma_0/4$ and $\gamma=4\gamma_0$ have been considered in the funnel simulations. The resulting powders piles showed a good agreement in terms of AOR values and powder surface morphology with funnel experiments considering more coarse-grained / fine-grained powders with increased / decreased mean particle diameters $\bar{d}=2\bar{d}_0$ and $\bar{d}=\bar{d}_0/2$. This observation confirmed the initial hypothesis that increasing / decreasing the mean powder particle size by a certain factor has an equivalent effect on the (quasi-static) bulk powder behavior as decreasing / increasing the surface energy by the square of this factor. Specifically, the choice $\gamma=4\gamma_0$, representing the case of very fine-grained powders with mean diameter $\bar{d}=14\mu m$, led to adhesive forces that dominate gravity forces by almost two orders of magnitude. In this regime, strongly increased AOR values as well as the presence of large cohesive particle agglomerates could be observed for the resulting powder piles.\\

A comparison with the (theoretical) case $\gamma=0$ revealed that a neglect of cohesive forces in the powder model leads to a drastic underestimation of the AOR and, consequently, to an insufficient description of the bulk powder behavior when employing the model in potential applications such as AM powder recoating simulations~\cite{Meier2018b}. Eventually, the present study demonstrated that the resulting angle of repose is more sensitive concerning the magnitude of surface energy as compared to the magnitude of other powder material parameters such as particle stiffness, friction coefficient or coefficient of restitution. The high sensitivity of the bulk powder behavior concerning the magnitude of surface energy as well as the strong variation of effective surface energy values for a given material pairing suggest to calibrate this model parameter on the basis of bulk powder experiments and simulations, while typical literature values can be taken for the remaining (less sensitive) material parameters in good approximation.\\

Research work going on in parallel focuses on the verification of the fitted surface energy values on particle level based on atomic force microscope (AFM) experiments, and on the application of the proposed powder model for analyzing the powder recoating process in metal additive manufacturing processes~\cite{Meier2018b}.

\section*{Acknowledgements}
\label{sec:Acknowledgements}

The authors would like to thank Otis R. Walton for fruitful discussions and valuable literature recommendations. Financial support for preparation of this article was provided by the German Academic Exchange Service (DAAD) to C. Meier and R. Weissbach as well as by Honeywell Federal Manufacturing \& Technologies to A.J. Hart. This work has been partially funded by Honeywell Federal Manufacturing \& Technologies, LLC which manages and operates the Department of Energy's Kansas City National Security Campus under Contract No. DE-NA-0002839.  The United States Government retains and the publisher, by accepting the article for publication, acknowledges that the United States Government retains a nonexclusive, paid up, irrevocable, world-wide license to publish or reproduce the published form of this manuscript, or allow others to do so, for the United States Government purposes. Moreover, this material is based upon work supported by the Assistant Secretary of Defense for Research and Engineering under Air Force Contract No. FA8721-05-C-0002 and/or FA8702-15-D-0001. Any opinions, findings, conclusions or recommendations expressed in this material are those of the author(s) and do not necessarily reflect the views of the Assistant Secretary of Defense for Research and Engineering.

%
\bibliographystyle{plain}
\bibliography{powdercoating_1.bib,review_heat_transfer_Jan2018_2.bib,beamreferences.bib}
%
%
\end{document}